We extend Merrifield's Variational Ansatz in the variational band theory of polarons to cover a frame of two electronic bands mixed by an Einstein phonon. The Hamiltonian is composed of the local and hopping energy terms, the vibrational energy, and a band-mixing term linear in the electron-phonon coupling. The eigenstate is a linear combination of Merrifield states for either electron band. The variational equations are solved numerically, so as to obtain the energy vs. momentum relation in ground state. Our variational method generates either Jahn-Teller polarons if the electronic bands are degenerate or Pseudo-Jahn-Teller polarons if they are nearly degenerate, both entities regarded as likely carriers in metal-oxygen manifolds of high-$T_c$ superconducting cuprates and colossal magnetoresistance exhibiting manganates.

Introduction. The variational Ansatz is a powerful tool for studying a polaron (charge carrier + associated distortion). It defines a specific variational eigenstate comprising fermion and boson ladder operators to diagonalize the Hamiltonian and deduce a minimal (ground state) energy. For a review of variational work see.[1] Aimed originally at Holstein polarons which form through the interaction of an electronic carrier with symmetry-retaining phonons along a linear chain, it should be extendable to lower symmetries by incorporating a vibronic mixing.

A variational Ansatz has earlier been applied to describing an itinerant lower-symmetry Jahn-Teller (JT) polaron in 1D.[2] The coupling term is assumed of the band-diagonal type in which the phonon field couples to the difference in electron density between the two electronic bands. However, this type of interaction Hamiltonian failing to provide any genuine mixing interband transitions, it is discarded by other authors as being nonvibronic.[3] Nevertheless, an energy vs. total momentum $E(\kappa)$ relation has been computed and found polaronic.

Our Merrifield-based analysis indicates that for band-diagonal coupling to the difference in electron density the system collapses to a single-band Holstein polaron. There appear to be two ways of splitting the electronic degeneracy in collective JT phenomena: (i) Genuine vibronic (band off-diagonal) coupling to a symmetry-breaking mode; (ii) Non-mixing (band-diagonal) coupling to a symmetry-retaining mode. The energy bands split in the latter case because the displaced-oscillator energy adds up differently. The lack of any genuine study of the vibronic polaron thus far puts an even stronger impetus on our work.

There has been an increasing appreciation lately of the possible role of vibronic polarons in the electric transport of transition-metal compounds.[4,5] The search for ε-θ JT polarons in $La_{2-x}Sr_xCuO_4$ has led to the discovery of a high-$T_c$ superconductivity. Similar ε–θ JT polarons are currently considered relative to the observed colossal magnetoresistance in $La_{1-x}Ca_xMnO_3$ and related materials. Pseudo-Jahn-Teller (PJT) polarons forming as fermionic excitations scatter from their associated double wells are mentioned as principal charge carriers maintaining the axial charge leak for an interplane coupling in single-layer cuprates, such as $La_{2-x}Sr_xCuO_4$.[6]

Hamiltonian. The vibronic polaron requires the availability of two narrow (nearly-) degenerate electronic bands and a mixing phonon field of the appropriate symmetry.[7] In so far as the band off-diagonal coupling is expected to break the original site symmetry, the mixing mode has to be a symmetry-breaking vibration transforming according to a irreducible representation of the point group. If the latter group contains the spatial inversion, the mixing mode will be odd parity if the respective electron bands compose of opposite-parity states. For instance, if the constituent bands form of 3d and 4p axial orbitals, even- and odd-parity, respectively, the mixing vibration may be one of the $A_{2u}$ or the $E_u$ odd-parity modes, both being symmetry-breaking in nature. For sufficiently strong coupling, the octahedral (tetragonal) site symmetry will be lowered to that of an $A_{2u}$- or $E_u$- deformed octahedron. Alternatively, the mixing mode will be even parity, such as the $E_g$ modes, if the respective electronic bands are degenerate. Now, the symmetry will be lowered to that of an $E_g$ deformed octahedron, e.g. tetragonal.

The vibronic Hamiltonian reads in second quantization terms:

$$H_{vib} \equiv H_{loc} + H_{kin} + H_{int} + H_{ph} =$$

$$\sum_{n,\mu} \varepsilon_{n,\mu} a_{n,\mu}^+ a_{n,\mu} + \sum_{n,\mu} j_{n\mu} a_{n,\mu}^+ (a_{n+1,\mu} + a_{n-1,\mu}) + \sum_{n,\mu,\nu} g^{n\mu\nu} a_{n,\mu}^+ a_{n,\nu} (b_n^+ + b_n) + \sum_n \hbar\omega_n b_n^+ b_n,$$

where $H_{loc}$, $H_{kin}$, $H_{int}$, and $H_{ph}$ denote the four sums in the order of their appearance: the local energy, the kinetic (hopping) energy, the electron-phonon interaction energy, and the phonon energy parts, respectively. Also n is the site label, $\mu$ and $\nu$ are band labels ($\mu,\nu=$ 1,2). $a_{n,\mu}^+$ ($a_{n,\mu}$) are fermion creation (annihilation) operators, $b_n^+$ ($b_n$) are boson creation (annihilation) operators, $\varepsilon_{n,\mu}$ are the local fermion energies, $j_{n,\mu}$ are the fermion hopping energies, $g^{n,\mu\nu}$ are the fermion-boson coupling constants, $\omega_n$ are the phonon frequencies. Point-group symmetry requires that the representation of the phonon field be included in the direct product of the electronic band representations: $\Gamma_{ph} \subset \Gamma_1 \otimes \Gamma_2$. Throughout, $h$ is Planck's constant / $2\pi$.

Variational Eigenstate. We build a variational eigenstate as a linear combination of Merrifield states for either of the constituent bands (q-phonon momentum, $\kappa$- total momentum):[8,9]

$$|\psi(\kappa)\rangle = \sum_\mu \alpha_\mu^\kappa |\psi_\mu(\kappa)\rangle$$

$$|\psi_\mu(\kappa)\rangle = N^{-1/2} \sum_n e^{i\kappa n} a_{n\mu}^+ \exp\{-N^{-1/2} \sum_q (\beta_{q\mu}^\kappa e^{-iqn} b_q^+ - \beta_{q\mu}^{\kappa *} e^{+iqn} b_q)\} |0\rangle$$

Here and above $\alpha_\mu$ stand for the fractional band amplitudes. The single-band Merrifield eigenstates normalize automatically. The normalization condition for their linear combination reads:

$$\langle\psi(\kappa)|\psi(\kappa)\rangle \equiv \sum_\mu |\alpha_\mu^\kappa|^2 = 1$$

Variational Equations. The present variational problem is one at the following parameters: $\beta_{q1}^\kappa$, $\beta_{q2}^\kappa$, $\alpha_1^\kappa$, and $\alpha_2^\kappa$, each set taken at a specific value of the total crystalline momentum $\kappa$. A system of self-consistent equations for $\mu,\nu=1,2$ is derived (details to appear elsewhere):[10]

$$\beta_{q\mu}^\kappa = \{(g^{\mu\mu}/\hbar\omega) + (g^{\mu\nu}/\hbar\omega)(\alpha_\nu^\kappa/\alpha_\mu^\kappa)S_{\nu\mu}^\kappa\}\{-D_{q\nu\mu}^\kappa + (g^{\nu\mu}/\hbar\omega)(\alpha_\mu^\kappa/\alpha_\nu^\kappa)S_{\mu\nu}^\kappa Q_{\nu\mu}^\kappa\}/D_q$$

$$\beta_{q\nu}^\kappa = \{(g^{\nu\nu}/\hbar\omega) + (g^{\nu\mu}/\hbar\omega)(\alpha_\mu^\kappa/\alpha_\nu^\kappa)S_{\mu\nu}^\kappa\}\{-D_{q\mu\nu}^\kappa + (g^{\mu\nu}/\hbar\omega)(\alpha_\nu^\kappa/\alpha_\mu^\kappa)S_{\nu\mu}^\kappa Q_{\mu\nu}^\kappa\}/D_q$$

$$|\alpha_\mu^\kappa| = \{1 + |\varepsilon_{\nu\mu}^\kappa|^2 / |E(\kappa) - \varepsilon_{\nu\nu}^\kappa|^2\}^{-1/2}$$

$$|\alpha_\nu^\kappa| = \{1 + |\varepsilon_{\mu\nu}^\kappa|^2 / |E(\kappa) - \varepsilon_{\nu\nu}^\kappa|^2\}^{-1/2}$$

where:

$$D_{q\mu\nu}^\kappa = 1 + 4(j_\mu/\hbar\omega) S_{\mu\mu}^\kappa \sin(\kappa - \Phi_{\mu\mu}^\kappa - q/2) \sin(q/2) - (g^{\mu\nu}/\hbar\omega)\text{Re}[(\alpha_\nu^\kappa/\alpha_\mu^\kappa) Q_{\nu\mu}^\kappa S_{\nu\mu}^\kappa]$$

$$D_q = D_{q\mu\nu}^\kappa D_{q\nu\mu}^\kappa - (g^{\mu\nu}/\hbar\omega)^2 |S_{\mu\nu}^\kappa|^2 |Q_{\mu\nu}^\kappa|^2$$

$$S_{\mu\mu}^\kappa = \exp\{-(1/N)\sum_q |\beta_{q\mu}^\kappa|^2 [1 - \cos(q)]\}$$

$$\Phi_{\mu\mu}^{\kappa} = (1/N) \sum_q |\beta_{q\mu}^{\kappa}|^2 \sin(q)$$
$$S_{\mu\nu}^{\kappa} = \exp\{-(1/2N)\sum_q(|\beta_{q\mu}|^2 + |\beta_{q\nu}|^2 - 2\beta_{q\mu}^{\kappa*}\beta_{q\nu}^{\kappa})\}$$
$$Q_{\mu\nu}^{\kappa} = (1/N)\sum_q(\beta_{-q\mu}^{\kappa*} + \beta_{q\nu}^{\kappa})$$
$$n_{\mu}^{\kappa} = (1/N) \sum_q |\beta_{q\mu}^{\kappa}|^2$$
$$\varepsilon_{\mu\mu}^{\kappa} = \varepsilon_{\mu} + 2 j_{\mu} S_{\mu\mu}^{\kappa} \cos(\kappa - \Phi_{\mu\mu}^{\kappa}) + g^{\mu\mu} Q_{\mu\mu}^{\kappa} + \hbar\omega n_{\mu}^{\kappa}$$
$$\varepsilon_{\mu\nu}^{\kappa} = g^{\mu\nu} Q_{\mu\nu}^{\kappa} S_{\mu\nu}^{\kappa}$$
$$E(\kappa) = \tfrac{1}{2} \{(\varepsilon_{\mu\mu}^{\kappa} + \varepsilon_{\nu\nu}^{\kappa}) \pm \sqrt{[(\varepsilon_{\mu\mu}^{\kappa} - \varepsilon_{\nu\nu}^{\kappa})^2 + 4 \varepsilon_{\mu\nu}^{\kappa} \varepsilon_{\nu\mu}^{\kappa}]}\}$$

$E(\kappa)$ is the extremal energy signifying the energy in the lowest polaron band.

Numerical Calculations. We made use of three languages: Basic, Pascal, and Fortran 77. Our present calculations confined to real numbers. An iterative program was worked out under appropriate starting conditions for the vibrational amplitudes in the form of a "small-polaron" distribution $\beta_{q\mu}^{\kappa}(0) = $ const in q-space. The latter was used to compute zeroth-order "starting values" for the Debye-Waller factors $S_{\mu\mu}^{\kappa}(0)$, $S_{\nu\nu}^{\kappa}(0)$, $S_{\mu\nu}^{\kappa}(0)$, phases $\Phi_{\mu\mu}^{\kappa}(0)$, $\Phi_{\nu\nu}^{\kappa}(0)$, mode coordinates $Q_{\mu\nu}^{\kappa}(0)$, and phonon occupation numbers $n_{\mu}^{\kappa}(0)$. Regarded as constants, these were inserted into the variational equations to derive first-order vibrational amplitudes $\beta_{q\mu}^{\kappa}(I)$, etc. thereby ending up the first iterative step. The procedure was further repeated with $\beta_{q\mu}^{\kappa}(I)$ used for computing parameters $S_{\mu\mu}^{\kappa}(I)$, etc. Ultimately $\beta_{q\mu}^{\kappa}(II)$ resulted as the latter were inserted into the variational equations to solve for the first-order phonon amplitudes. Such iteration steps were taken $N' = 20\text{-}30$ times until the band amplitudes $\alpha_{\mu}^{\kappa}(N')$ met the normalization condition to a reasonable accuracy (better than 4%).

Vibronic Polarons. Introducing definitions in concert with the traditional regime of parameters used for polaron research, we divide the electron energy axis into three ranges as follows:
- (I) Low-energy range where $|e_{\mu\nu}| / \hbar\omega \ll 1$,
- (II) Intermediate-energy range where $|e_{\mu\nu}| / \hbar\omega \sim 1$,
- (III) High-energy range where $|e_{\mu\nu}| / \hbar\omega \gg 1$

where $|e_{\mu\nu}| = |\varepsilon_{\mu} - \varepsilon_{\nu}|$ is the nearly-degenerate states' gap energy but it can also be formulated in terms of any other energy-dependent electronic parameter, such as the coupling constant $g^{\mu\nu}$ and the hopping energy $j_{\mu\mu}$. It is motivated by the adiabatic approximation being expandable in powers of $\hbar\omega / |e_{\mu\nu}|$ which enters as a "small quantity". Range III is traditional for polaron study by means of the adiabatic approximation. Ranges I and II have not been explored so far but doubts have been raised as to the applicability of adiabatic methods to them.[11]

With reference to range I, we distinguish the following polaron types depending on the relative values of the remaining parameters at $G_{\mu\nu} \neq 0$, all electron parameters now expressed in units of a phonon quantum $G_{\mu\nu} = g^{\mu\nu} / \hbar\omega$, $J_{\mu} = j_{\mu} / \hbar\omega$, $E_{\mu\nu} = |e_{\mu\nu}| / \hbar\omega$:
- (i) Adiabatic polaron (AD) for $G_{\mu\nu}^2 < 2 J_{\mu}$;
- (ii) Antiadiabatic polaron (AAD) for $G_{\mu\nu}^2 > 2 J_{\mu}$;
- (iii) Jahn-Teller polaron (JT) for $E_{\mu\nu} = 0$;
- (iv) Pseudo-Jahn-Teller polaron (PJT) for $E_{\mu\nu} \neq 0$;
- (v) Weakly coupled polaron (WC) for $2G_{\mu\nu}^2 < E_{\mu\nu}$;
- (vi) Strongly coupled polaron (SC) for $2G_{\mu\nu}^2 > E_{\mu\nu}$;
- (vii) Dicoupled polaron (DC) for $G_{\mu\mu} = 0$, $G_{\nu\nu} \neq 0$;
- (viii) Tricoupled polaron (TC) for $G_{\mu\mu}, G_{\nu\nu} \neq 0$;
- (ix) Semibound polaron (SB) for $J_{\mu} = 0$, $J_{\nu} \neq 0$;
- (x) Bound polaron (B) for $J_{\mu} = J_{\nu} = 0$.

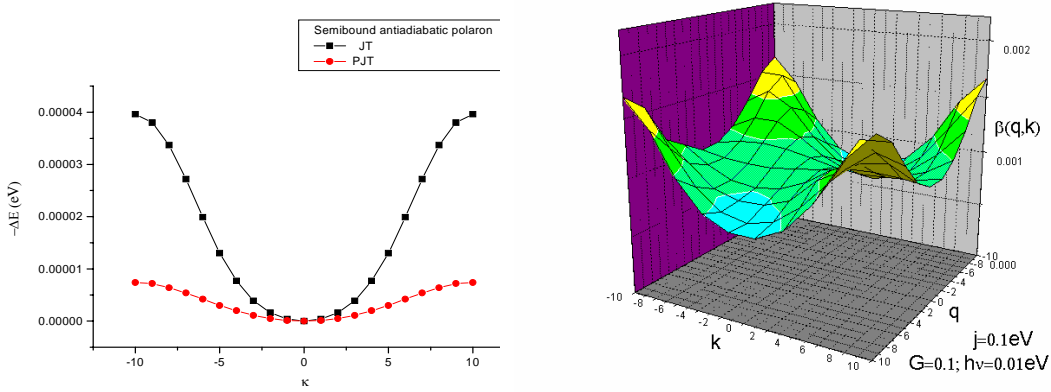

Figure 1. Calculated E(κ) of vibronic polarons and β(q,κ) of Merrifield-Holstein polarons

We made a number of calculations on these polaron types in the nonadiabatic Range I. To simulate situations closer to actual β-distributions, one-band Merrifield equaions were used for generating starting conditions in the form of Holstein polarons for either component band. The band amplitudes started with $\alpha_\mu^\kappa(0) = 1/\sqrt{2}$. All the calculations were made at a phonon energy of 0.01 eV at total momentum κ in units of π/10. The vibronic variational equations generated small polarons (with phonon amplitudes $\beta_{q\mu}^\kappa$ independent of the phonon momentum q).

Conclusion. This investigation was intended to starting a series of variational calculations on vibronic polarons. Range I was worth studying, since it overlaps with an essential portion of the "nonadiabaticity range" for which little is known, in so far as the adiabatic approximation, traditional source of analytic conclusions, does not hold true therein. We believe the present investigation does have a heuristic value in raising a few points of scientific importance.

The phonon dressing effect is demonstrated to be slight in Range I. The vibronic variational equations generate small polarons, strongly confined in real space. This numerical result is not surprising in view of the expected size of configurational distortion, arising from the local character of the band mixing by phonons. Another intriguing observation is that the polaron effect increases as it should from adiabatic to antiadiabatic and from JT to PJT. We also consider it untrivial that stable semibound vibronic polarons generate which contributes to our understanding of the behavior of carriers trapped in ground electronic state which can migrate if in the excited electronic state: the ground state-excited state mixing makes them all itinerant. Stable dicoupled and tricoupled polarons are prohibited by group theory on grounds of the incompatible mode symmetries in that $g^{\mu\nu}$ and $g^{\mu\mu}$ may not be both finite for a given mode symmetry. Yet, we see these polarons firmly itinerant raising the belief that certain selection-breaking compromise may eventually be found in experimentally important cases. Range I variational conclusions for bound polarons may not be compared against the background of analytic data, since the adiabatic approximation does not apply to gap energies < $h\omega$.

Further work is planned to incorporate the complex plane so as to extend the regime to larger values of the vibronic parameters, possibly to Range III. This would make it possible to compare variational results with adiabatic results. A missing analytic result is the transition from small to large bound PJT polarons, as the enegy gap $E_{\mu\nu}$ increases towards the critical value of $2(G^{\mu\nu})^2$ at constant $G^{\mu\nu}$. This textbook conclusion has been drawn for the adiabatic energy Range III. Another result is the unadiabatic behavior of the mixing-mode coordinate $Q_{\mu\nu}$ of bound JT polarons found to increase as the gap increased, in contrast to the Range III result. It is perhaps not surprising now why our Range I solutions yield small polarons only. Linking Range I to Range III is of chief importance for understanding the vibronic polarons.

References.